\documentclass[12pt]{article}

\usepackage[dvips]{graphicx}

\def\dspace{\baselineskip = 0.30in}

\def\lapproxeq{\lower .7ex\hbox{$\;\stackrel{\textstyle
<}{\sim}\;$}}
\def\gapproxeq{\lower .7ex\hbox{$\;\stackrel{\textstyle
>}{\sim}\;$}}

\begin{document}

\dspace

\begin{titlepage}
\begin{flushright}
%\preprint{
BA-02-36\\
%March 2002\\
%}
\end{flushright}
\vskip 2cm
\begin{center}
%\title
{\Large\bf
GUT Scale And Leptogenesis From 5D Inflation
%5D Inflation, Leptogenesis and the Supersymmetric Grand Unification Scale
}
\vskip 1cm
{\normalsize\bf
%\author{
Bumseok Kyae\footnote{bkyae@bartol.udel.edu} and
Qaisar Shafi\footnote{shafi@bxclu.bartol.udel.edu}
}
\vskip 0.5cm
{\it Bartol Research Institute, University of Delaware, \\Newark,
DE~~19716,~~USA\\[0.1truecm]}

%
%%\maketitle

\end{center}
\vskip .5cm

%\date{\today}
%\pacs{PACS: 11.25.Mj, 12.10.Dm, 98.80.Cq}

\begin{abstract}

We discuss a five dimensional inflationary scenario based on
a supersymmetric $SO(10)$ model compactified on $S^1/(Z_2\times Z_2')$.
Inflation is implemented through scalar potentials
on four dimensional branes, and a brane-localized Einstein-Hilbert term
is essential to make both brane vacuum energies positive
during inflation.
The orbifold boundary conditions break the $SO(10)$ gauge symmetry
to $SU(4)_c\times SU(2)_L\times SU(2)_R$ ($\equiv H$).
The inflationary scenario yields $\delta T/T\propto (M/M_{\rm Planck})^2$,
which fixes $M$, the symmetry breaking scale of $H$ to be close
to the SUSY GUT scale of $10^{16}$ GeV.
The scalar spectral index $n$ is $0.98-0.99$, while the gravitational wave
contribution to the quadrupole anisotropy is negligible ($\lapproxeq 1 {\%}$).
The inflaton decay into the lightest right handed neutrinos
yields the observed baryon asymmetry via leptogenesis.

\end{abstract}
\end{titlepage}

\newpage

%%%%%%%%%
%%%%%%%%%

%%%%%%%%%%%%%%%%%%%%%%%%%%%%%%%%%%%%%%%%%%%%%%%%%%%%%%%%%%%%%%%%%%%%%%%%%%%
%\section{Introduction}

There exists a class of supersymmetric models in which a close link
exists between inflation and
the grand unification scale~\cite{hybrid, review}.
In particular, the quadrupole microwave anisotropy is proportional
to $(M/M_ {\rm Planck})^2$,
where $M$ denotes the scale of the gauge symmetry breaking
associated with inflation, and $M_{\rm Planck}=1.2\times 10^{19}$ GeV.
Thus, $M$ is expected to be of order $10^{16}$ GeV,
to within a factor of 2 or so, depending on the details of
the supersymmetric model. This is tantalizingly close to the
supersymmetric grand unification scale inferred from the evolution of
the minimal supersymmetric standard model (MSSM) gauge couplings,
and it is therefore natural to try to realize this inflationary scenario
within a grand unified framework~\cite{review}.
The $SO(10)$ model is particularly attractive in view of the growing confidence
in the existence of neutrino oscillations~\cite{nuoscil},
which require that at least two of the three known neutrinos have a non-zero
mass. Because of the presence of right handed  neutrinos (MSSM singlets),
non-zero masses for the known neutrinos is an automatic consequence
of the see-saw mechanism~\cite{seesaw}.
Furthermore, the right handed neutrinos play an essential role in generating
the observed baryon asymmetry via leptogenesis~\cite{lepto},
which becomes especially compelling within an inflationary
framework~\cite{ls}.
Indeed, an inflationary scenario would be incomplete
without explaining the origin of the observed baryon asymmetry,
and the kind of models we are interested in here
automatically achieve this via leptogenesis.

A realistic supersymmetric inflationary model along
the lines we are after was presented in \cite{khalil},
based on the $SO(10)$ subgroup
$SU(4)_c\times SU(2)_L\times SU(2)_R$ ($\equiv H$)~\cite{ps}.
The scalar spectral index $n$ has a value very close to unity
(typically $n\approx 0.98-0.99$),
while the symmetry breaking scale of $H$ lies,
as previously indicated, around $10^{16}$ GeV.
The vacuum energy density during inflation is of order $10^{14}$ GeV,
so that the gravitational contribution to the quadrupole anisotropy
is essentially negligible.  It is important to note here that
the inflaton field in this scenario eventually decays into right handed
neutrinos, whose out of equilibrium decays lead to leptogenesis.
An extension to the full $SO(10)$ model is complicated by the notorious
doublet-triplet splitting problem,
which prevents a straightforward implementation of the inflationary scenario.
Of course, the subgroup $H$ neatly evades this problem and even allows for
a rather straightforward resolution of the `$\mu$ problem.'

Our objective here is to take advantage of recent orbifold constructions
of five dimensional (5D) supersymmetric GUTs, in which a grand unified symmetry
such as $SO(10)$ can be readily broken to its maximal subgroup
$H$~\cite{dermisek} (an alternative possiblity is $SU(5)\times U(1)$
which we will not pursue here), with the doublet-triplet splitting problem
circumvented without fine tuning of parameters.
Our main challenge then is to develop a 5D framework which can be merged
with the four dimensional (4D) supersymmetric inflationary scenario
based on $H$.
Because of $N=2$ SUSY (in 4D sense) in 5D bulk, the F-term inflaton
potential is allowed only on the 4D orbifold fixed points (branes),
where only $N=1$ SUSY is preserved.
We shall see how 4D inflation comes about
through scalar potentials localized on the two branes
by analyzing the 5D Einstein equation.
A brane-localized Einstein-Hilbert term is essential
to make both brane vacuum energies positive definite during inflation,
which is a condition required by 4D $N=1$ SUSY.

The four dimensional inflationary model is best illustrated by considering
the following superpotential which allows the breaking of some gauge
symmetry $G$ down to $SU(3)_c\times SU(2)_L\times U(1)_Y$,
keeping supersymmetry (SUSY) intact \cite{hybrid, lyth}:
\begin{eqnarray} \label{simplepot}
W_{\rm infl}=\kappa S(\phi\bar{\phi}-M^2) ~.
\end{eqnarray}
Here $\phi$ and $\bar{\phi}$ represent superfields
whose scalar components acquire non-zero vacuum expectation values (VEVs).
For the particular example of $G=H$ above,
they belong to the ${\bf (\overline{4}, 1,2)}$ and ${\bf (4,1,2)}$
representations of $H$.
The $\phi$, $\bar{\phi}$ VEVs break $H$ to the MSSM gauge group.
The singlet superfield $S$ provides the scalar field that drives inflation.
Note that by invoking a suitable $R$ symmetry $U(1)_R$,
the form of $W$ is unique at the renormalizable level, and it
is gratifying to realize that $R$ symmetries naturally occur
in (higher dimensional) supersymmetric theories
and can be appropriately exploited.
From $W$, it is straightforward to show that the supersymmetric minimum
corresponds to non-zero (and equal in magnitude) VEVs
for $\phi$ and $\bar{\phi}$, while $\langle S\rangle =0$~\cite{king}.
(After SUSY breaking {\it $\grave{a}$ la} $N=1$ supergravity
(SUGRA), $\langle S\rangle$ acquires a VEV
of order $m_{3/2}$ (gravitino mass)).

An inflationary scenario is realized in the early universe
with both $\phi$, $\bar{\phi}$ and $S$ displaced
from their present day minima.
Thus, for $S$ values in excess of the symmetry breaking scale $M$,
the fields $\phi$, $\bar{\phi}$ both vanish,
the gauge symmetry is restored, and a potential energy density proportional
to $M^4$ dominates the universe. With SUSY thus broken, there are
radiative corrections from the $\phi$-$\bar{\phi}$ supermultiplets
that provide logarithmic corrections to the potential which drives inflation.
In one loop approximation \cite{hybrid, coleman},
\begin{eqnarray}\label{scalarpot}
V\approx V_{\rm tree}+\kappa^2M^4\frac{\kappa^2N}{32\pi^2}\bigg[
2{\rm ln}\frac{\kappa^2|S|^2}{\Lambda^2}+(z+1)^2{\rm ln}(1+z^{-1})
+(z-1)^2{\rm ln}(1-z^{-1})\bigg]~,
\end{eqnarray}
where $z=x^2=|S|^2/M^2$, $\Lambda$ denotes a renormalization mass scale and
$N$ denotes the dimensionality of the $\phi$, $\bar{\phi}$ representations.
From Eq.~(\ref{scalarpot}) the quadrupole anisotropy is
found to be~\cite{hybrid, review}
\begin{eqnarray}\label{T}
\bigg(\frac{\delta T}{T}\bigg)_Q\approx \frac{8\pi}{\sqrt{N}}
\bigg(\frac{N_Q}{45}\bigg)^{1/2}\bigg(\frac{M}{M_{\rm Planck}}\bigg)^2
x_Q^{-1}y_Q^{-1}f(x_Q^2)^{-1} ~.
\end{eqnarray}
The subscript $Q$ is there to emphasize the epoch of horizon crossing,
$y_Q\approx x_Q(1-7/12x_Q^2+\cdots)$, $f(x_Q^2)^{-1}\approx 1/x_Q^2$,
for $S_Q$ sufficiently larger than $M$,
and $N_Q\approx 45-60$ denotes the e-foldings needed to resolve the horizon
and flatness problems.
From the expression for $\delta T/T$ in Eq.~(\ref{T}) and comparison with the
COBE result $(\delta T/T)_Q\approx 6.6\times 10^{-6}$~\cite{cobe},
it follows that the gauge symmetry breaking scale $M$ is
close to $10^{16}$ GeV.  Note that $M$ is associated in our $SO(10)$ example
with the breaking scale of $H$ (in particular the $B-L$ breaking scale),
which need not exactly coincide with the SUSY GUT scale.
We will be more specific about $M$ later in the text.

The relative flatness of the potential ensures that the primordial density
fluctuations are essentially scale invariant.
Thus, the scalar spectral index $n$ is
0.98 for the simplest example based on $W$ in Eq.~(\ref{simplepot}).
In some models $n$ is unity to within a percent.

Several comments are in order:

\noindent $\bullet$ The 50-60 e-foldings required to solve the horizon and
flatness problems occur when the inflaton field $S$ is relatively close
(to within a factor of order 1-10) to the GUT scale.
Thus, Planck scale corrections can be safely ignored.

\noindent $\bullet$ For the case of minimal K${\rm\ddot{a}}$hler potential,
the SUGRA corrections do not affect the scenario at all,
which is a non-trivial result~\cite{review}.
More often than not, supersymmetric inflationary scenarios fail to work
in the presence of SUGRA corrections which tend to spoil the flatness
of the potential needed to realize inflation.

\noindent $\bullet$ Turning to the subgroup $H$ of $SO(10)$,
one needs to take into account the fact that the spontaneous breaking of $H$
produces magnetic monopoles that carry two quanta of
Dirac magnetic charge~\cite{magg}.
An overproduction of these monopoles at or near
the end of inflation is easily avoided, say by introducing an additional
(non-renormalizable) term $S(\phi\bar{\phi})^2$ in $W$,
which is permitted by the $U(1)_R$ symmetry.
The presence of this term ensures the absence of monopoles as explained
in Ref.~\cite{khalil}. Note that the monopole problem is also
avoided by choosing a different subgroup of $SO(10)$.
In a separate publication, we will consider a scenario based on the
$SU(3)_c\times SU(2)_L\times U(1)_Y\times U(1)_X$ subgroup of $SO(10)$
whose breaking does not lead to monopoles.
Another interesting candidate is
$SU(3)_c\times SU(2)_L\times SU(2)_R\times U(1)_{B-L}$.
The salient features of the model are not affected by the monopole
problem~\cite{khalil}.

\noindent $\bullet$ At the end of inflation the scalar fields
$\phi$, $\bar{\phi}$, and $S$ oscillate about their respective minima.
Since the $\phi$, $\bar{\phi}$ belong respectively
to the ${\bf (\overline{4},1,2)}$
and ${\bf (4,1,2)}$ of $SU(4)_c\times SU(2)_L\times SU(2)_R$,
they decay exclusively into right handed neutrinos
via the superpotential couplings,
\begin{eqnarray} \label{nu}
W=\frac{\gamma_{i}}{M_P}\bar{\phi}\bar{\phi}F^c_iF^c_i ~,
\end{eqnarray}
where the matter superfields $F^c_i$ belong to
the ${\bf (\overline{4},1,2)}$ representation of $H$, and
$M_P\equiv M_{\rm Planck}/\sqrt{8\pi}=2.44\times 10^{18}$ GeV denotes
the reduced Planck mass, and $\gamma_{i}$ are dimensionless coefficients.
We will have more to say about inflaton decay,
the reheat temperature, as well as leptogenesis taking account of the recent
neutrino oscillation data.
However, we first wish to provide a five dimensional
setting for this inflationary scenario.

We consider 5D space-time ($x^\mu, y$),
$\mu=0,1,2,3$, where the fifth dimension is compactified on an $S^1/Z_2$
%(or $S^1/(Z_2\times Z_2')$)
orbifold.
The action is given by
\begin{eqnarray} \label{action}
S=\int d^4x \int_{-y_c}^{y_c}dy\sqrt{-g_5}\bigg[\frac{M_5^3}{2}~R_5-\Lambda_B
+\frac{\delta(y)}{\sqrt{g_{55}}}\bigg(\frac{M_4^2}{2}~\bar{R}_4-\Lambda_1\bigg)
-\frac{\delta(y-y_c)}{\sqrt{g_{55}}}\Lambda_2\bigg] ~,
\end{eqnarray}
where $R_5$ ($\bar{R}_4$) is the 5 dimensional (4 dimensional)
Einstein-Hilbert term\footnote{The importance of the
brane-localized 4D Einstein-Hilbert term, especially for generating 4D gravity
with a non-compact extra dimension was first noted
in Ref.~\cite{braneR}.},
$\Lambda_B$, and $\Lambda_1$, $\Lambda_2$ are the bulk and
brane cosmological constants, and $M_5$ and $M_4$ are mass parameters.
The cosmological constants in the bulk and on the branes could be interpreted
the vacuum expectation values of some scalar potentials
from the particle physics sector.
The brane curvature scalar (Ricci scalar) $\bar{R}_4(\bar{g}_{\mu\nu})$
is defined with the induced metric of the bulk metric,
$\bar{g}_{\mu\nu}(x)\equiv g_{\mu\nu}(x,y=0)$
($\mu,\nu=0,1,2,3$).
For an inflationary solution, we take the metric ansatz,
\begin{eqnarray} \label{metric}
ds^2=\beta^2(y)(-dt^2+e^{2H_0t}d\vec{x}^2)+dy^2 ~,
\end{eqnarray}
where $H_0$ could be interpreted as the 4 dimensional Hubble constant.
The non-vanishing components $(\mu,\mu)$ and $(5,5)$
of the 5 dimensional Einstein equation
derived from (\ref{action}) gives \cite{nihei}
\begin{eqnarray} \label{eom}
&&3\bigg[\bigg(\frac{\beta'}{\beta}\bigg)^2+\bigg(\frac{\beta''}{\beta}\bigg)
-\bigg(\frac{H_0}{\beta}\bigg)^2
-\delta(y)\frac{M_4^2}{M_5^3}\bigg(\frac{H_0}{\beta}\bigg)^2\bigg] \\
&&~~~~~~~~~~~=-\frac{\Lambda_B}{M_5^3}-\delta(y)\frac{\Lambda_1}{M_5^3}
-\delta(y-y_c)\frac{\Lambda_2}{M_5^3} ~,  \nonumber \\
&&6\bigg[\bigg(\frac{\beta'}{\beta}\bigg)^2-\bigg(\frac{H_0}{\beta}\bigg)^2
\bigg]=-\frac{\Lambda_B}{M_5^3} ~, \label{eom2}
\end{eqnarray}
where primes denote derivatives with respect to $y$.
The last term in the left hand side in Eq.~(\ref{eom})
arises from the brane scalar curvature term, and vanishes when $H_0=0$.

The solutions to the equations Eq.~(\ref{eom}) and (\ref{eom2})are given by
\begin{eqnarray} \label{sol1}
\beta(y)&=&\bigg(\frac{H_0}{k}\bigg)~{\rm sinh}(\pm k|y|+c)
~~~{\rm for}~~~\Lambda_B<0 ~,  \\
\beta(y)&=&\pm H_0|y|+c ~~~~~~~~~~~~~~~~~~{\rm for}~~~\Lambda_B=0  ~,
\label{sol2}
\end{eqnarray}
where $k\equiv \sqrt{-\Lambda_B/6M_5^3}$, and $c$ is an integration constant.
Without loss of generality, we can take $c$ positive.
To avoid the existence of naked singularities within the interval
$-y_c<y<y_c$, $\pm ky_c+c>0$ and $\pm H_0y_c+c>0$ should be required.
For simplicity of our discussion, let us take `$+$' among $\pm$
in Eqs.~(\ref{sol1}) and (\ref{sol2}).

The introduction of the brane scalar curvature term $\bar{R}_4$
does not affect the bulk solutions, (\ref{sol1}) and (\ref{sol2}),
but it modifies the boundary conditions.
For $\Lambda_B<0$, the solution should satisfy
the following boundary conditions at $y=0$ and $y=y_c$,
\begin{eqnarray} \label{bdy1}
&&k~{\rm coth}c-\frac{1}{2}\frac{M_4^2}{M_5^3}~\frac{k^2}{{\rm sinh}^2c}
=-\frac{\Lambda_1}{6M_5^3}  ~,  \\
&&k~{\rm coth}(ky_c+c)=\frac{\Lambda_2}{6M_5^3} ~.  \label{bdy2}
\end{eqnarray}
Hence, the integration constant $c$ and the interval length $y_c$
are determined by $\Lambda_1$ and $\Lambda_2$.
Similarly, the solution for $\Lambda_B=0$ should satisfy
the boundary conditions,
\begin{eqnarray} \label{bdy1'}
&&\frac{H_0}{c}-\frac{1}{2}\frac{M_4^2}{M_5^3}~\frac{H_0^2}{c^2}
=-\frac{\Lambda_1}{6M_5^3} ~, \\
&&\frac{H_0}{c+H_0y_c}=\frac{\Lambda_2}{6M_5^3} ~, \label{bdy2'}
\end{eqnarray}
so $\Lambda_1$ and $\Lambda_2$ determine $H_0/c$ and $y_c$.
Note that $\Lambda_2$ must be fine-tuned to zero
when $\Lambda_1=0$~\cite{selftun}.
Hence it is natural that the scalar field which controls inflation
is introduced in the bulk.

From Eqs.~(\ref{bdy1})--(\ref{bdy2'}), we note that the brane cosmological
constants $\Lambda_1$ and $\Lambda_2$ should have opposite signs
in the absence of the brane curvature scalar contribution at $y=0$.
However, a suitably large value of $M_4/M_5$ can even make
the sign of $\Lambda_1$ positive.
Since the introduction of the brane curvature term does not conflict with
any symmetry that may be present,
there is no reason why such a term with a parameter $M_4$ that is large
compared to $M_5$ is not allowed \cite{braneR}.
Thus, $\Lambda_1$ and $\Lambda_2$ could both be positive
and this fact will be exploited for implementing the inflationary scenario.
We will later suggest a model for explaining how a large $M_4/M_5$ ratio
may be realized.

From (\ref{bdy1'}) and (\ref{bdy2'}), we also note that
$\Lambda_1$ and $\Lambda_2$ in the $\Lambda_B=0$ case are related
to the 4 dimensional Hubble constant $H_0$,
unlike the $\Lambda_B<0$ case in Eqs.~(\ref{bdy1}) and (\ref{bdy2}).
While their non-zero values are responsible for the 3-space
inflation, vanishing brane cosmological constants guarantee an effective
4 dimensional flat space-time.
On the other hand, for $\Lambda_B<0$,
the relations between the bulk and brane cosmological
constants are responsible for inflation.
To obtain a static solution \cite{rs},
we should take $H_0\rightarrow 0$ and $c\rightarrow\infty$
(or $\Lambda_{1(2)}/6M_5^3\rightarrow -k~(+k)$) while letting the ratio
$H_0e^c/2k\rightarrow 1$.

Our main task is to embed the 4D supersymmetric inflationary scenario
in 5D space-time~\cite{king2},
employing the framework and solutions discussed above.
In order to extend the setup to 5D SUGRA, a gravitino $\psi_M$
and a vector field $B_M$ should be appended
to the graviton (f${\rm\ddot{u}}$nfbein) $e_{M}^{m}$.
Through orbifolding, only $N=1$ SUSY is preserved on the branes.
The brane-localized Einstein-Hilbert term in Eq.~(\ref{action})
is still allowed, but should
be accompanied by a brane gravitino kinetic term as well as other terms,
which is clear in off-shell SUGRA formalism \cite{kyae}.
In a higher dimensional supersymmetric theory, a F-term scalar potential is
allowed only on the 4 dimensional fixed points
which preserve $N=1$ SUSY.
We require a formalism in which inflation and
the Hubble constant $H_0$ are controlled
only by the brane cosmological constants, such that
during inflation the positive vacuum energy slowly decreases, and
the minimum of the scalar potential corresponds to a flat 4D space-time.
The solution for $\Lambda_B=0$ meets these requirements
in the presence of the additional brane scalar curvature term at $y=0$,
and so we will focus only on this case.

We have tacitly assumed that the interval separating the two branes
(orbifold fixed points) remains fixed during inflation.
The scenario is quite different from what is often called
`D-brane inflation' \cite{Dbrane}.
The dynamics of the orbifold fixed points, unlike the D-brane case,
is governed only by the $g_{55}(x,y)$ component of the metric tensor.
The real fields $e_5^5$, $B_5$, and the chiral fermion $\psi^2_{5R}$
in the 5D gravity multiplet are assigned even parity under $Z_2$ \cite{kyae},
and they compose an $N=1$ chiral multiplet on the branes.
The associated superfield can acquire a superheavy mass and
its scalar component can develop a VEV on the brane.
With superheavy brane-localized mass terms,
their low-lying Kaluza-Klein (KK) mass spectrum is shifted
so that even the lightest mode obtains
a compactification scale mass~\cite{localmass}.
Since this is much greater than $H_0$, the interval distance is stable
even during inflation.
The stabilization of the interval distance leads to the stabilization also
of the warp factor $\beta(y)$, because the fluctuation $\delta \beta(y)$
of the warp factor near the solution
in Eq.~(\ref{sol2}) (also Eq.~(\ref{sol1})) turns out to be proportional to
the interval length variation $\delta g_{55}$
by the linearized 5D Einstein equation~\cite{chacko}.

With $\Lambda_B=0$, the effective 4 dimensional reduced Planck mass squared
$M_P^2~(\equiv 1/8\pi G_N)$ is given by
\begin{eqnarray}
M_{P}^2&\equiv& M_5^3\int_{-y_c}^{y_c}dy\beta^2+M_4^2\beta^2|_{y=0} ~,
\nonumber \\
&=&M_5^3y_c\bigg[\frac{2}{3}H_0^2y_c^2+2cH_0y_c+2c^2\bigg]+M_4^2c^2 ~.
\end{eqnarray}
%
%which is the coefficient of the effective 4 dimensional Einstein-Hilbert term,
%and define the Planck mass.
%
For $M_5^3y_c<<M_4^2\sim M_P^2$, gravity couples universally at low energy
to fields localized at $y=y_c$ and $y=0$ and in the bulk,
with the strength controlled by $1/M_4^2$ \cite{kyae2}.
The 4 dimensional effective cosmological constant turns out to be
\begin{eqnarray}
\Lambda_{\rm eff}&=&\int_{-y_c}^{y_c}dy\beta^4\bigg[M_5^3\bigg(
4\bigg(\frac{\beta''}{\beta}\bigg)+6\bigg(\frac{\beta'}{\beta}\bigg)^2\bigg)
+\delta(y)\Lambda_1+\delta(y-y_c)\Lambda_2\bigg]  \nonumber \\
&=&3H_0^2~\bigg[M_5^3y_c~\bigg(\frac{2}{3}H_0^2y_c^2+2cH_0y_c
+2c^2\bigg)+M_4^2c^2\bigg] ~=~ 3H_0^2M_P^2 ~,
\end{eqnarray}
where the first two terms in the first line are the warp factor contributions.
Hence, from Eqs.~(\ref{bdy1'}) and (\ref{bdy2'}) $\Lambda_{\rm eff}$ vanishes
when $\Lambda_1=\Lambda_2=0$.  Note that for $H_0y_c<<1$,
\begin{eqnarray} \label{4dcc}
\Lambda_{\rm eff}\approx \frac{c^2}{12}~\frac{\Lambda_2^2M_P^2}{M_5^6} ~.
\end{eqnarray}
We can directly adapt these results for the $S^1/(Z_2\times Z_2')$ case.

To see how inflation is realized in this 5D setting,
let us consider the 4D $SU(4)_c\times SU(2)_L\times SU(2)_R (\equiv H)$
supersymmetric inflationary model \cite{khalil}.
An effective 4D theory with the gauge group $H$ is readily obtained
from a 5D $SO(10)$ gauge theory if the fifth dimension is compactified
on the orbifold $S^1/(Z_2\times Z_2')$ \cite{dermisek}, where
$Z_2$ reflects $y\rightarrow -y$, and $Z_2'$ reflects $y'\rightarrow -y'$
with $y'=y+y_c/2$.
There are two independent orbifold fixed points (branes)
at $y=0$ and $y=y_c/2$, with $N=1$ SUSYs and gauge symmetries $SO(10)$
and $H$ respectively \cite{dermisek}.
The $SO(10)$ gauge multiplet $(A_M,\lambda^1,\lambda^2,\Phi)$
decomposes under $H$ as
\begin{eqnarray}
V_{\bf 45}&\longrightarrow& V_{({\bf 15},{\bf 1},{\bf 1})}+
V_{({\bf 1},{\bf 3},{\bf 1})}+V_{({\bf 1},{\bf 1},{\bf 3})}
+V_{({\bf 6},{\bf 2},{\bf 2})} \\
&&+\Sigma_{({\bf 15},{\bf 1},{\bf 1})}
+\Sigma_{({\bf 1},{\bf 3},{\bf 1})}+\Sigma_{({\bf 1},{\bf 1},{\bf 3})}
+\Sigma_{({\bf 6},{\bf 2},{\bf 2})} ~,  \nonumber
\end{eqnarray}
where $V$ and $\Sigma$ denote the vector multiplet $(A_\mu,\lambda^1)$ and
the chiral multiplet $((\Phi+iA_5)/\sqrt{2},\lambda^2)$ respectively,
and their $(Z_2,Z_2')$ parity assignments and KK masses
are shown in Table I.
\vskip 0.6cm
\begin{center}
\begin{tabular}{|c||c|c|c|c|} \hline
Vector &$V_{({\bf 15},{\bf 1},{\bf 1})}$ & $V_{({\bf 1},{\bf 3},{\bf 1})}$ &
$V_{({\bf 1},{\bf 1},{\bf 3})}$ &$V_{({\bf 6},{\bf 2},{\bf 2})}$ \\
\hline \hline
$(Z_2,Z_2')$ &(+,+)$$ &$(+,+)$ &$(+,+)$ &$(+,-)$ \\
Masses & $2n\pi/y_c$ & $2n\pi/y_c$ & $2n\pi/y_c$ & $(2n+1)\pi/y_c$ \\
\hline \hline
Chiral & $\Sigma_{({\bf 15},{\bf 1},{\bf 1})}$ &
$\Sigma_{({\bf 1},{\bf 3},{\bf 1})}$ &
$\Sigma_{({\bf 1},{\bf 1},{\bf 3})}$ &$\Sigma_{({\bf 6},{\bf 2},{\bf 2})}$
\\ \hline \hline
$(Z_2,Z_2')$ & $(-,-)$ &$(-,-)$ &$(-,-)$ &$(-,+)$  \\
Masses & $(2n+2)\pi/y_c$ & $(2n+2)\pi/y_c$ & $(2n+2)\pi/y_c$ & $(2n+1)\pi/y_c$
\\
\hline
\end{tabular}
%\vskip 0.2cm
\end{center}
{\bf Table I.~}($Z_2,Z_2'$) parity assignments and
Kaluza-Klein masses $(n=0,1,2,\cdots)$ ~~~~~~~~~~~~
\hspace{2.0cm} for the vector multiplet in $SO(10)$.
% \end{center}
\vskip 0.2cm

The parities of the chiral multiplets $\Sigma$'s are opposite to those of
the vector multiplets $V$'s in Table I and
hence, $N=2$ SUSY explicitly breaks to $N=1$
below the compactification scale $\pi/y_c$.
As shown in Table I, only the vector multiplets,
$V_{({\bf 15},{\bf 1},{\bf 1})}$, $V_{({\bf 1},{\bf 3},{\bf 1})}$, and
$V_{({\bf 1},{\bf 1},{\bf 3})}$ contain massless modes, which
means that the low energy effective 4D theory reduces to $N=1$
supersymmetric $SU(4)_c\times SU(2)_L\times SU(2)_R$.
The parity assignments in Table I also show that the wave function
of the vector multiplet $V_{({\bf 6},{\bf 2},{\bf 2})}$ vanishes
at the brane located at $y=y_c/2$ (B2) because it is assigned an odd parity
under $Z_2'$, while the wave functions of all the vector multiplets should
be the same at the $y=0$ brane (B1).
Therefore, while the gauge symmetry at B1 is $SO(10)$,
only $SU(4)_c\times SU(2)_L\times SU(2)_R$ is preserved at B2 \cite{hebecker}.

The 5D inflationary solution requires positive vacuum energies on both branes
B1 and B2.  While the scalar potential in Eq.~(\ref{scalarpot})
would be suitable for B2, an appropriate scalar potential on B1 is
also required.
Since the boundary conditions in Eq.~(\ref{bdy1'}) and (\ref{bdy2'})
require $\Lambda_1$ and $\Lambda_2$ to simultaneously vanish,
it is natural to require $S$ to be a bulk field.
Then, the VEVs of $S$ on the two branes can be adjusted
such that the boundary conditions are satisfied.
As an example, consider the following superpotential on B1,
\begin{eqnarray} \label{b1superpot}
W_{B1}=\kappa_1S(Z\overline{Z}-M_1^2) ~,
\end{eqnarray}
where $Z$ and $\overline{Z}$ are $SO(10)$ singlet superfields on the B1 brane
with opposite $U(1)_R$ charges.
The condition for a positive brane cosmological constant
on B1 is found from (\ref{bdy1'}) to be $(H_0/c)(M_4^2/M_5^3)>2$.
For $\kappa\sim 10^{-3}$, say, and $c\sim 1$, we have
$H_0\sim 10^{10}$ GeV and $M_5\sim 10^{15}$ GeV (so that $M_4\sim M_P$).
Thus, there exists a hierarchy of order $10^{3}$
between the 5D bulk scale $M_5$ and
the four dimensional brane mass scale $M_4$.
To see how this hierarchy could arise, consider the case where
the brane-localized gravity kinetic term has the canonical form
but not the bulk term.  Thus,
\begin{eqnarray}
{\cal L}/e=
\frac{M_P^3}{2}e^{-f(|\phi|)}R_5
+\frac{\delta(y)}{e_5^5}\bigg(\frac{M_P^2}{2}~\bar{R}_4-V(|\phi|)\bigg)
-\frac{1}{2y_c}\partial_M\phi\partial^M\phi^*
%-\delta(y-y_c)V_2(\phi)
+\cdots ~,
%\nonumber
\end{eqnarray}
where $\phi$ is some scalar field, $V(|\phi|)$ its associated potential,
we take $M_4=M_P$.
Let us assume that like $e_5^5$, $\phi$ acquires
a Planck scale mass and VEV on the brane.
Then, at the minimum of $V(|\phi|)$,
%below the Planck scale $\phi$ is decoupled and
the 5D Einstein equation determining the background geometry
is effectively given by Eqs.~(\ref{eom}) and (\ref{eom2}),
with $M_5=e^{-\langle f(|\phi|)\rangle/3}M_P$.
Taking $f(|\phi|)=2|\phi|/M_P$, for instance, and
$\langle\phi\rangle\approx 10M_P$
would lead to $M_5\sim 10^{15}$ GeV as required.

After inflation is over, the oscillating system consists of the complex scalar
fields $\Phi=(\delta\bar{\phi}+\delta\phi)$,
where $\delta\bar{\phi}=\bar{\phi}-M$ ($\delta \phi=\phi-M$), and $S$,
both with masses equal to $m_{\rm infl}=\sqrt{2}\kappa M$.
Through the superpotential couplings in Eq.~(\ref{nu}),
these fields decay into a pair of right handed neutrinos and sneutrinos
respectively, with an approximate decay width \cite{khalil}
\begin{eqnarray}\label{decay}
\Gamma\sim \frac{m_{\rm infl}}{8\pi}\bigg(\frac{M_i}{M}\bigg)^2 ~,
\end{eqnarray}
where $M_i$ denotes the mass of the heaviest right handed neutrino
with $2M_i<m_{\rm infl}$, so that the inflaton decay is possible.
Assuming an MSSM spectrum below the GUT scale, the reheat temperature
is given by \cite{hybrid2}
\begin{eqnarray} \label{temp}
T_r\approx \frac{1}{3}\sqrt{\Gamma M_P}
\approx\frac{1}{12}\bigg(\frac{55}{N_Q}\bigg)^{1/4}\sqrt{y_Q}M_i ~.
\end{eqnarray}
For $y_Q\sim$ unity (see below), and $T_r\lapproxeq 10^{9.5}$ GeV
from the gravitino constraint~\cite{gravitino},
we require $M_i\lapproxeq 10^{10}-10^{10.5}$ GeV.

In order to decide on which $M_i$ is involved in the decay~\cite{pati},
let us start with atmospheric neutrino ($\nu_\mu-\nu_\tau$) oscillations and
assume that the light neutrinos exhibit an hierarchical mass pattern
with $m_3>>m_2>>m_1$.
Then $\sqrt{\Delta m^2_{\rm atm}}\approx m_3\approx m_{D3}^2/M_3$,
where $m_{D3}$ ($=m_t(M)$)
denotes the third family Dirac mass which equals the asymptotic top quark
mass due to $SU(4)_c$.  We also assume a mass hierarchy in the right handed
sector, $M_3>>M_2>>M_1$.
The mass $M_3$ arises from the superpotential coupling Eq.~(\ref{nu})
and is given by
$M_3= 2\gamma_{3}M^2/M_P\sim 10^{14}~~{\rm GeV}$,
for $M\sim 10^{16}$ GeV and $\gamma_{3}\sim$ unity.  This value of $M_3$ is in
the right ball park to generate an $m_3\sim \frac{1}{20}$ eV
($\sim \sqrt{\Delta m_{\rm atm}^2}$),
with $m_t(M)\sim 110$ GeV \cite{hybrid2}.
It follows from (\ref{temp}) that $M_i$ in (\ref{decay}) cannot be identified
with the third family right handed neutrino mass $M_3$.
It should also not correspond to the second family neutrino mass $M_2$
if we make the plausible assumption that the second generation Dirac mass
should lie in the few GeV scale.
The large mixing angle MSW solution of the solar neutrino
problem requires that
$\sqrt{\Delta m_{\rm solar}^2}\approx m_2\sim {\rm GeV}^2/M_2
\sim \frac{1}{160}~{\rm eV}$,
so that $M_2\gapproxeq 10^{11}-10^{12}$ GeV.
Thus, we are led to conclude~\cite{pati} that
the inflaton decays into the lightest (first family) right handed neutrino
with mass
\begin{eqnarray}\label{M1}
M_1\sim 10^{10}-10^{10.5} ~{\rm GeV} ~,
\end{eqnarray}
such that $2M_1<m_{\rm infl}$.

The constraint $2M_2> m_{\rm infl}$ yields
$y_Q\lapproxeq 3.34 \gamma_{2}$,
where $M_2=2\gamma_{2}M^2/M_P$.  We will not provide here a comprehensive
analysis of the allowed parameter space but will be content to present
a specific example, namely
\begin{eqnarray}\label{value}
M\approx 8\times 10^{15}~{\rm GeV}~,~~\kappa\approx 10^{-3}~,~~
m_{\rm infl}\sim 10^{13}~{\rm GeV}~(\sim M_2)~,
\end{eqnarray}
with $y_Q\approx 0.4$ (corresponding to $x_Q$ near unity, so that the inflaton
$S$ is quite close to $M$ during the last 50--60 e-foldings).

Note that typically $\kappa$ is of order
$10^{-2}$-- few $\times 10^{-4}$~\cite{khalil},
so that the vacuum energy density during inflation is
$\sim 10^{-4}-10^{-8}~M_{\rm GUT}^4$.
Thus, in this class of models the gravitational wave contribution to
the quadrupole anisotropy $(\delta T/T)_Q$ is essentially negligible
($\lapproxeq 10^{-8}$).
With $\kappa\sim{\rm few}\times 10^{-4}$ ($10^{-3}$),
the scalar spectral index $n\approx 0.99$ ($0.98$).

The decay of the (lightest) right handed neutrinos generates
a lepton asymmetry which is given by \cite{lasymm}
\begin{eqnarray} \label{lasymm}
\frac{n_L}{s}\approx \frac{10}{16\pi}\bigg(\frac{T_r}{m_{\rm infl}}\bigg)
\bigg(\frac{M_1}{M_2}\bigg)
\frac{c_{\theta}^2s_{\theta}^2~{\rm sin}2\delta~(m_{D2}^2-m_{D1}^2)^2}
{|\langle h\rangle|^2(m_{D2}^2s_{\theta}^2+m_{D1}^2c_{\theta}^2)} ~,
\end{eqnarray}
where the VEV $|\langle h\rangle|\approx 174$ GeV (for large ${\rm tan}\beta$),
$m_{D1,2}$ are the neutrino Dirac masses (in a basis in which they are
diagonal and positive), and $c_\theta\equiv {\rm cos}\theta$,
$s_\theta\equiv {\rm sin}\theta$, with $\theta$ and $\delta$ being the rotation
angle and phase which diagonalize the Majorana mass matrix of
the right handed neutrinos.
Assuming $c_\theta$ and $s_\theta$ of comparable magnitude,
taking $m_{D2}>>m_{D1}$, and using (\ref{M1}) and (\ref{value}),
Eq.~(\ref{lasymm}) reduces to
\begin{eqnarray}
\frac{n_L}{s}\approx 10^{-8.5}c_\theta^2 ~{\rm sin}2\delta ~
\bigg(\frac{T_r}{10^{9.5}~{\rm GeV}}\bigg)
\bigg(\frac{M_1}{2\cdot 10^{10.5}~{\rm GeV}}\bigg)
\bigg(\frac{10^{13}~{\rm GeV}}{M_2}\bigg)
\bigg(\frac{m_{D2}}{10~{\rm GeV}}\bigg)^2~,
\end{eqnarray}
which can be in the correct ball park to account for the observed baryon
asymmetry $n_B/s$ ($\approx -28/79 ~n_L/s$).

In conclusion, our goal in this paper has been to demonstrate the existence
of realistic models which nicely blend together four particularly attractive
ideas, namely supersymmetric grand unification, extra dimension(s),
inflation and leptogenesis.
%
%Indeed, recent experimental measurements provide a huge boost for the
%inflationary scenario.
The doublet-triplet problem is circumvented by utilizing orbifold
breaking of $SO(10)$, which may also help in suppressing dimension five
proton decay.
There are two predictions concerning inflation that are particularly
significant.  Namely, the scalar spectral index $n$ lies very close to unity
($\approx 0.98-0.99$), and the gravitational wave contribution to
$(\delta T/T)_Q$ is highly suppressed ($\sim 10^{-8}-10^{-9}$).
Finally, the inflaton decay produces heavy right handed Majorana neutrinos
(in our case the lightest one), whose subsequent out of equilibrium decay
leads to the baryon asymmetry via leptogenesis.
We expect to generalize this approach to
other symmetry breaking patterns of $SO(10)$ in a future publication.

\vskip 0.3cm
\noindent {\bf Acknowledgments}

\noindent
Q.S. thanks Gia Dvali and George Lazarides for useful discussions.
We also acknowledge useful discussion with Jim Liu.
The work is partially supported
by DOE under contract number DE-FG02-91ER40626.

\end{document}